\documentclass[pdflatex,sn-mathphys-num]{sn-jnl}

\usepackage{graphicx}%
\usepackage{multirow}%
\usepackage{amsmath,amssymb,amsfonts}%
\usepackage{amsthm}%
\usepackage{mathrsfs}%
\usepackage[title]{appendix}%
\usepackage{xcolor}%
\usepackage{textcomp}%
\usepackage{manyfoot}%
\usepackage{booktabs}%
\usepackage{algorithm}%
\usepackage{algorithmicx}%
\usepackage{algpseudocode}%
\usepackage{listings}%
\usepackage{ulem}%
\usepackage{caption}
\usepackage{setspace}
\newcommand{\mycomment}[1]{}
\renewcommand{\appendixname}{Methods}
\newcommand{\MechMIT}{Department of Mechanical Engineering, Massachusetts Institute of Technology, 
    Cambridge, MA 02139, USA}

\begin{document}

\title[Article Title]{Conditional Motional Squeezing of an Optomechanical Oscillator Approaching the Quantum Regime}


\author[1]{\fnm{Benjamin} \spfx{B.} \sur{Lane}}\equalcont{These authors contributed equally to this work.}

\author[1]{\fnm{Junxin} \sur{Chen}}
\equalcont{These authors contributed equally to this work.}

\author[2]{\fnm{Ronald} \spfx{E.} \sur{Pagano}}

\author[2]{\fnm{Scott} \sur{Aronson}}

\author[3]{\fnm{Garrett} \spfx{D.} \sur{Cole}}

\author[1]{\fnm{Xinghui} \sur{Yin}}

\author[1,4]{\fnm{Vivishek} \sur{Sudhir}}

\author[2]{\fnm{Thomas} \spfx{R.} \sur{Corbitt}}

\author*[1]{\fnm{Nergis} \sur{Mavalvala}}\email{nergis@ligo.mit.edu}

\affil*[1]{\orgdiv{LIGO}, \orgname{Massachusetts Institute of Technology}, \orgaddress{\street{185 Albany Street}, \city{Cambridge}, \state{Massachusetts}, \postcode{02139}, \country{USA}}}

\affil[2]{\orgdiv{Department of Physics and Astronomy}, \orgname{Louisiana State University}, \orgaddress{\street{Tower Drive}, \city{Baton Rouge}, \state{Louisiana},
\postcode{70803}, \country{USA}}}

\affil[3]{\orgname{Thorlabs Crystalline Solutions}, \orgaddress{\street{114 E Haley St., Suite G}, \city{Santa Barbara}, \state{California},
\postcode{93101}, \country{USA}}}

\affil[4]{\MechMIT}


\abstract{Squeezed mechanical states are a highly-coveted resource for quantum-enhanced sensing and serve as a compelling platform for probing the interplay between gravity and quantum mechanics. It has been predicted that a mechanical oscillator can be prepared into a quantum squeezed state if the applied measurement rate is fast relative to its mechanical resonance frequency. However, the experimental feasibility of this protocol has remained uncertain because of the difficulty in achieving low-frequency oscillators with sufficiently strong read-out. Here, we demonstrate that a careful selection of parameters in an optomechanical system, combined with optimal filtering techniques, enables the preparation of a $50$~ng GaAs cantilever in a conditional classical squeezed state, achieving a minimum uncertainty of just $1.07\pm0.04$ times the zero-point fluctuation level. This minimum variance is $3$ orders of magnitude smaller than what has been achieved in previous experiments using the same protocol. Although we do not fully achieve the quantum squeezed regime, our demonstration provides definitive evidence that a measurement-based protocol is a practical and effective approach for the real-time preparation of macroscopic oscillators in quantum squeezed states. }

\keywords{macroscopic quantum state preparation, quantum optomechanics, quantum state estimation}



\maketitle

%\section{Introduction}\label{sec:intro}

Preparing and controlling quantum states in mechanical systems at mesoscopic and macroscopic scales has become a frontier in exploring the nature of the quantum-to-classical transition, which seeks to understand how classical behavior emerges from quantum laws \cite{Diosi1984,penrose1996gravity,Bassi2003Dynamical,chen2013macroscopic}. Investigating this transition involves probing various hypothetical decoherence mechanisms proportional to the mass of the quantum state. 
Moreover, preparing and stabilizing mesoscopic to macroscopic systems in quantum states provide a potential testbed for studying the quantum nature of gravity, where macroscopic superpositions and entanglement generated by gravity could offer insights into whether gravity itself is quantum \cite{dewitt2011role,carney2019tabletop}. Recently, schemes employing Gaussian states have been proposed as promising alternatives with reduced experimental demands.\cite{Ludovico2024Testing,Kryhin2025Distinguishable}.
Mechanical oscillators in their quantum ground states - where fluctuations in displacement and momentum quadratures saturate the Heisenberg uncertainty relation - are a fundamental goal in the field of quantum state preparation \cite{Aspelmeyer2014Cavity}. 
In the ground state, energy is equally distributed between displacement and momentum quadratures, referred to as equipartition of energy.

Beyond this equipartitioned state lies the quantum squeezed state, where noise in one quadrature can be reduced below the zero-point fluctuation (ZPF) at the cost of increasing noise in the conjugate quadrature. 
As a result, squeezed states exhibit significantly greater sensitivity to changes in the noise-reduced quadrature than the ground state. This enhanced sensitivity makes them valuable for technologies that require ultraprecise measurements.  \cite{hollenhorst1979quantum,Aspelmeyer2014Cavity,lo2015spin}

The key to preparing mechanical quantum squeezed states is to break the equipartition of energy between the displacement $\hat{q}$ and the momentum $\hat{p}$ quadratures. 
To date, solid-state mechanical oscillators have been prepared in quantum squeezed states using reservoir engineering on microwave devices \cite{wollman2015quantum,pirkkalainen2015squeezing,lecocq2015quantum,Lei2016Quantum}. 
Furthermore, non-Gaussian quantum squeezed states have been successfully prepared by introducing coupling to superconducting qubits \cite{ma2021non,marti2024quantum}. 
In the optical domain, the approach analogous to reservoir engineering has successfully achieved back-action evasion but falls short by several orders of magnitude in the achievable squeezing due to excess noise, such as optical heating \cite{shomroni2019optical}. 
In contrast, inspired by ion trapping techniques \cite{Xin2021Rapid}, where quantum-squeezed mechanical states are prepared by modulating the trapping potential, quantum squeezing of the center-of-mass motion of a particle levitated in laser fields has been both inferred (after readout noise subtraction \cite{rossi2024quantum}) and directly measured \cite{kamba2025quantum}. 
So far, all successful demonstrations of mechanical quantum-squeezed state preparation have relied on parametric interactions. 

Another avenue for preparing mechanical quantum states is continuous measurement, which offers simpler and steadier experimental control. 

Feedback cooling of mechanical oscillators to the ground state using measurement rates faster than the thermal decoherence rate has already been achieved \cite{rossi2018measurement,tebbenjohanns2021quantum,magrini2021real}. 
However, in these demonstrations, energy equipartition between the displacement and momentum quadratures is maintained because mechanical oscillations are much faster than the measurement rates, preventing the emergence of observable squeezing.
In this regime, the rotating wave approximation is valid, implying that the dynamics do not support the preparation of squeezed states.

More than two decades ago, it was predicted that a measurement rate $\Gamma_\mathrm{meas}$ comparable to or exceeding the mechanical resonance frequency $\Omega_\mathrm{m}$ can violate energy equipartition and enable the preparation of quantum-squeezed states \cite{Doherty1999Feedback}. 
This approach has been refined over the last decades by incorporating state estimation and conditional state preparation \cite{Ebhardt2009Quantum,Meng2020Mechanical}.

\section*{Measurement-Based Preparation of a Squeezed Mechanical State}

State estimation is a widely used technique for extracting physical observables from noisy measurements and constitutes a central element  in this work.
A more detailed introduction can be found in Methods~\ref{app:StateEst}.
For any linear continuous measurement of displacement $\hat{q}$, the measurement outcome can be normalized to displacement units as $\hat{y}(t)=\hat{q}(t)+\hat{z}(t)$, where $\hat{z}(t)$ is the normalized measurement imprecision noise.
To extract $\hat{q}(t)$ from a noisy measurement record $\hat{y}(t)$, an estimation filter $H_q(t)$ can be constructed and applied using a priori knowledge of the system, such as system parameters and noise spectra \cite{Kailath1981,Wieczorek2015Optimal,Rossi2019Observing}. 
The displacement estimator is subsequently given by $\hat{q}_\mathrm{e}(t)=H_q*\hat{y}(t)$, with $*$ denoting convolution.
The error of the estimate is given by $\delta\hat{q}(t)=\hat{q}(t)-\hat{q}_\mathrm{e}(t)$.
One can similarly define $\hat{p}_\mathrm{e}$ and $\delta\hat{p}$ for momentum $\hat{p}$. 

\begin{figure}[h]
    \centering
    \includegraphics[width=1\linewidth]{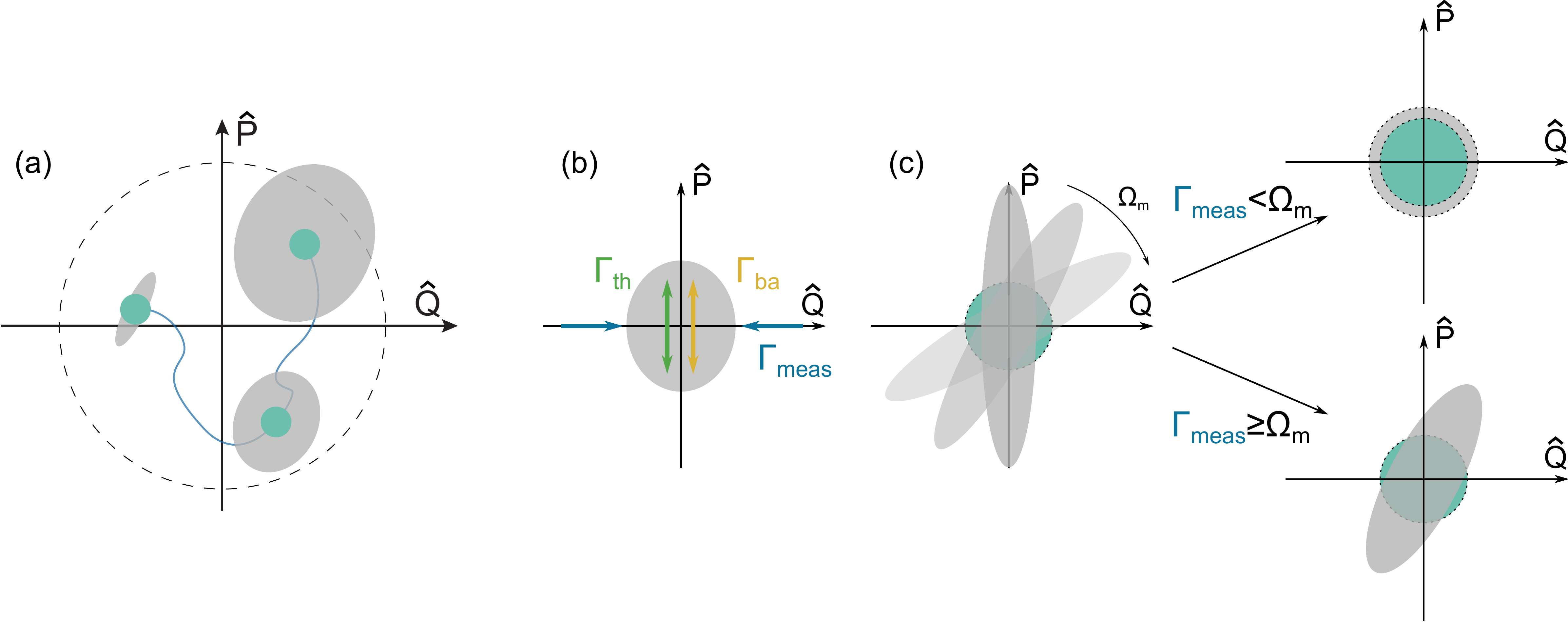}
    \caption{\textbf{(a)} Evolution of a conditional state. The axes use dimensionless units, defined as $\hat{Q} = \hat{q}/q_\mathrm{zpf}$ and $\hat{P} = \hat{p}/p_\mathrm{zpf}$, where the zero-point fluctuations are given by $q_\mathrm{zpf} = \sqrt{\hbar/2m\Omega_\mathrm{m}}$ and $p_\mathrm{zpf} = \sqrt{\hbar m\Omega_\mathrm{m}/2}$. The gray shaded area represents the estimation errors (i.e. the uncertainties of the conditional state), the green disk marks the magnitude of the ZPF, and the dashed circle represents the thermal distribution. The center of the conditional state undergoes a random walk, while the uncertainties decrease deterministically, converging to a steady state. \textbf{(b)} Uncertainties of the conditional state. The thermal decoherence rate $\Gamma_\mathrm{th}$ and back-action rate $\Gamma_\mathrm{ba}$ increase the uncertainty in the momentum quadrature, while the measurement rate $\Gamma_\mathrm{meas}$ decreases the uncertainty in the displacement quadrature. \textbf{(c)} The effect of quadrature rotation on the uncertainties of the conditional state. }
    \label{fig:intuition}
\end{figure}

For Gaussian states, the mean values of the estimators, $\langle \hat{q}_\mathrm{e} \rangle$ and $\langle \hat{p}_\mathrm{e} \rangle$, along with the (co)variances of their corresponding errors, $V_{\delta q \delta q}$, $V_{\delta p \delta p}$, and $C_{\delta q \delta p}$, uniquely define a state centered at the estimated mean values in phase space, with uncertainties determined by the estimation errors \cite{Ebhardt2009Quantum, Meng2020Mechanical, chen2024causal}.
Since estimators and their associated errors depend on the measurement record $\hat{y}$, the resulting state is termed a conditional state. The center of this conditional state undergoes a random walk in phase space, distributed according to a thermal state defined by thermal and back-action baths, as illustrated in Fig.~\ref{fig:intuition}(a). In contrast, the variances of the estimation errors decrease deterministically over time and converge to steady-state values as more information is acquired through measurement \cite{Rossi2019Observing}. A conditional state prepared using causal filters, which rely solely on past information, can be converted into an unconditional state through real-time feedback \cite{BoutHand08}. Consequently, causal conditional states are typically employed for state preparation.

In the following, we discuss how the rates and mechanical frequency influence the steady-state estimation errors. 
The Heisenberg-Langevin equations for a cavity optomechanical system are
\begin{align}
    \dot{\hat{X}}(t)&=-\frac{\kappa}{2}\hat{X}(t)-\Delta\hat{Y}(t)+\sqrt{\kappa}\hat{X}_\mathrm{in}(t)\\
    \dot{\hat{Y}}(t)&=-\frac{\kappa}{2}\hat{Y}(t)+\Delta\hat{X}(t)+\sqrt{2}G\alpha\hat{q}(t)+\sqrt{\kappa}\hat{Y}_\mathrm{in}(t)\label{eq:HLeqY}\\
    \dot{\hat{q}}(t)&=\frac{\hat{p}(t)}{m}\label{eq:HLeqq}\\
    \dot{\hat{p}}(t)&=-m\Omega_\mathrm{m}^2\hat{q}(t)+\sqrt{2}\hbar G\alpha\hat{X}(t)-\Gamma_\mathrm{m}\hat{p}(t)+\hat{\xi}_\mathrm{th}(t)\label{eq:HLeqp},
\end{align}
where $\hat{X}$ and $\hat{Y}$ are the amplitude and phase fluctuations of the intra-cavity optical field respectively; 
$\kappa$ is the angular cavity linewidth; $\Delta$ is the angular cavity detuning;
$\hat{X}_\mathrm{in}$ and $\hat{Y}_\mathrm{in}$ are vacuum amplitude and phase noises, respectively, coupled into the intra-cavity field. 
For simplicity, we do not distinguish between the different sources of vacuum noise. For a complete model, please see Methods \ref{app:OMT}. 
$G=\partial \Omega_\mathrm{cav}/\partial \hat{q}$ is the optomechanical interaction strength, with $\Omega_\mathrm{cav}$ the cavity angular resonance frequency;
$\alpha=\sqrt{n_\mathrm{cav}}$ is the classical amplitude of the intra-cavity optical field, with $n_\mathrm{cav}$ the intra-cavity photon number; 
$m$ is the effective mass of the oscillator;  %$G=\partial \Omega_\mathrm{cav}/\partial \hat{q}$ is the optomechanical interaction strength, with $\Omega_\mathrm{cav}$ the cavity angular resonance frequency;
%$\alpha=\sqrt{n_\mathrm{cav}}$ is the classical amplitude of the intra-cavity optical field, with $n_\mathrm{cav}$ the intra-cavity photon number; and  
%$\hat{X}$ is the amplitude fluctuation of the intra-cavity optical field; 
$\Gamma_\mathrm{m}$ is the natural angular mechanical linewidth; and $\hat{\xi}_\mathrm{th}$ is the noise due to the thermal force. 

The third term on the right-hand side of Eqn.~\ref{eq:HLeqY} encodes the mechanical displacement $\hat{q}$ into the optical field. 
Extracting this information via optical detection improves the estimation of $\hat{q}$ and thereby decreases the uncertainty $\delta\hat{q}$ at a rate of $\Gamma_\mathrm{meas}$, as shown in Fig.~\ref{fig:intuition}(b). 
$\Gamma_\mathrm{meas}$ is proportional to $n_\mathrm{cav}$, $\eta$, and $1/\kappa$, with $\eta$ the detection efficiency.  
At the same time, the second term on the right-hand side of Eq.~\ref{eq:HLeqp} is the back-action noise, which directly couples only to the momentum quadrature at a rate of $\Gamma_\mathrm{ba}$. 
Similarly, the thermal force noise couples directly to the momentum quadrature at a rate of $\Gamma_\mathrm{th}=\Gamma_\mathrm{m}n_\mathrm{th}$ with $n_\mathrm{th}$ the thermal phonon occupancy of the environment. 
These two noises tend to increase the uncertainty in $\delta\hat{p}$, as shown in Fig.~\ref{fig:intuition}(b). 
$\Gamma_\mathrm{meas}$, $\Gamma_\mathrm{ba}$ and $\Gamma_\mathrm{th}$ naturally tend to squeeze the mechanical state. 
However, the quadrature rotation described by Eq.~\ref{eq:HLeqq} indirectly couples the back-action noise and thermal force noise to $\hat{q}$, counteracting the uncertainty reduction achieved through measurement, as illustrated in Fig.~\ref{fig:intuition}(c). 
In most optomechanical systems, where $\Gamma_\mathrm{meas}\ll\Omega_\mathrm{m}$, the quadrature rotation averages out the squeezing induced by measurement, resulting in a circular (non-squeezed) conditional state, as shown in the upper branch of Fig.~\ref{fig:intuition}(c). 
If the measurement rate is comparable to or faster than the angular frequency of the mechanical oscillator, i.e. $\Gamma_\mathrm{meas}\gtrsim\Omega_\mathrm{m}$, back-action and thermal noises coupled to $\hat{p}$ are rotated to $\hat{q}$ slower than the reduction of $\delta\hat{q}$ by measurement, and equipartition can be violated. Thus the squeezing from measurement is preserved, as shown in the lower branch of Fig.~\ref{fig:intuition}(c). 

The condition $\Gamma_\mathrm{meas} \gtrsim \Omega_\mathrm{m}$ is most readily achieved with low-frequency mechanical oscillators. However, due to excess classical noise, previous experiments with such oscillators have been unable to prepare quantum-squeezed states. Instead, they produced states considered classically squeezed states that violate energy equipartition between $\hat{q}$ and $\hat{p}$ but do not exhibit quadrature uncertainties below the zero-point fluctuations (ZPF) \cite{santiago2020verification, meng2022measurement}.
Along the coveted path of quantum-squeezed motional states, the previous record set by \cite{santiago2020verification} achieved a standard deviation of the displacement quadrature that was 36 times ZPF, i.e. 1296 times the variance.

In this work, we demonstrate the preparation of a $50$~ng oscillator in a classically squeezed conditional mechanical state using causal Wiener filters in subsequent analysis. (For a linear system driven by stationary Gaussian noise, it is well known that the Wiener filter minimizes the mean square of estimation errors \cite{Kailath1981}.) 
The lowest achieved quadrature uncertainty in variance was merely $(0.28 \pm 0.18)$~dB above the ZPF, corresponding to a factor of $(1.07 \pm 0.04)$ times the ZPF, where the uncertainty reflects one standard deviation of the variance. The uncertainty in the conjugate quadrature was $(8.98 \pm 0.11)$~dB above the ZPF, or $(7.9 \pm 0.2)$ times of the ZPF, clearly demonstrating a violation of energy equipartition. This result represents a state-of-the-art achievement in measurement-based motional state squeezing of a mechanical oscillator, surpassing the previous record by three orders of magnitude. It marks a significant milestone toward reaching the quantum regime.
Our work shines light on the problem of the quantum-to-classical transition, especially the interplay between gravity and quantum states, by providing a heavy and easy-to-use sensor \cite{Diosi1984,penrose1996gravity,Ludovico2024Testing}. 
Due to the relatively large mass of our oscillator, it is also an ideal platform to demonstrate quantum-enhanced accelerometry \cite{krause2012high}. 
Moreover, complex optomechanical systems like LIGO are often plagued by instabilities that must be controlled. Our system includes a relatively simple control system, paving the way to preparing LIGO test masses in quantum-squeezed states without the risk of catastrophic instability.

\section*{Experimental Setup}\label{sec:experiment}

\begin{figure}[t]
    \centering
    \includegraphics[width=0.85\linewidth]{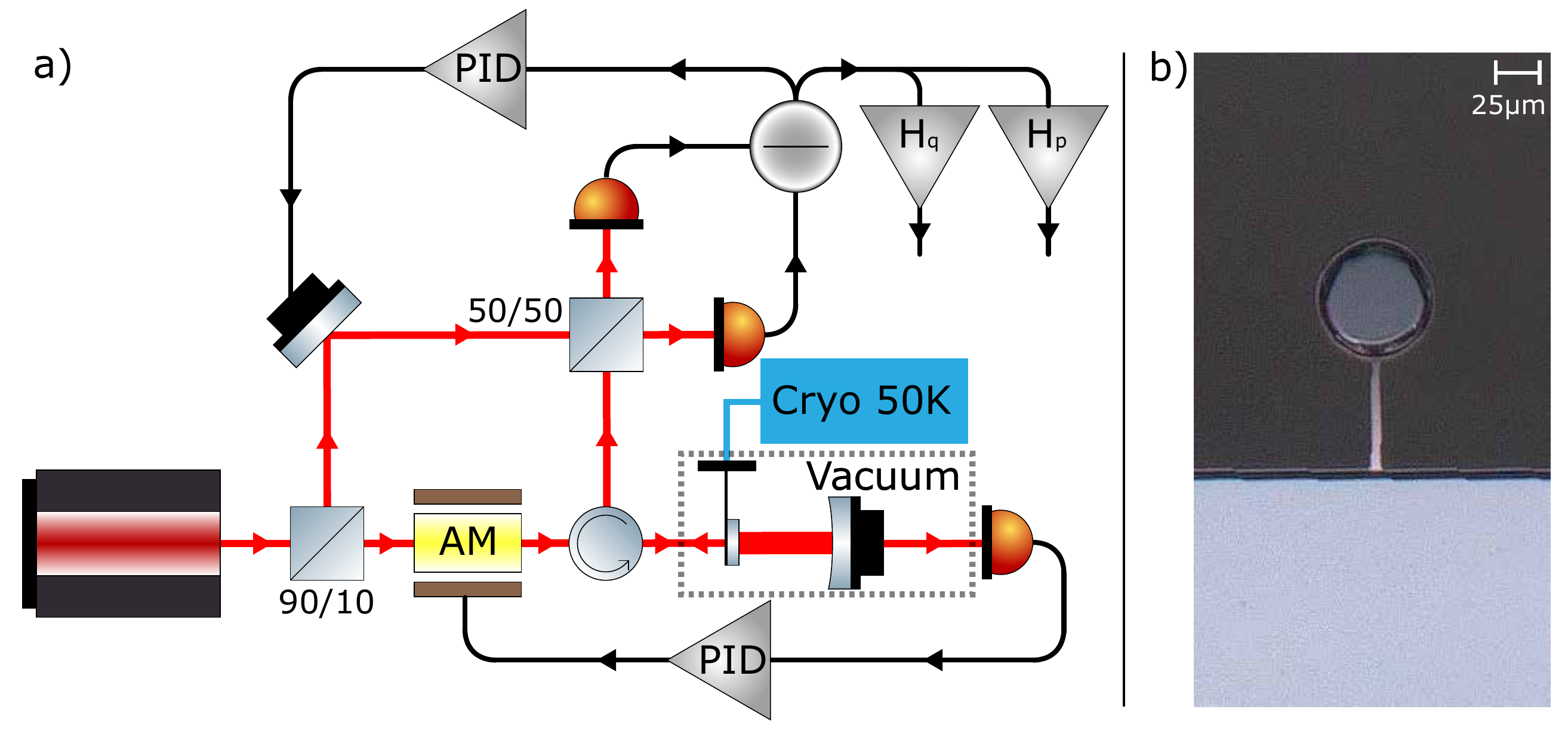}
    \caption{\textbf{(a)} Simplified optical layout of the experiment. AM, PID, $H_q$ and $H_p$ stand for optical amplitude modulator, PID servo, causal Wiener filter for $\hat{q}$ and $\hat{p}$ respectively. The input light is blue-detuned from the cavity resonance, stiffening the mechanical oscillator. The system dynamics are stabilized by feeding back to the amplitude of the input light. Displacement information is read out by a balanced homodyne receiver in reflection. The photocurrent is saved for subsequent analysis. \textbf{(b)} The optomechanical resonator as imaged with a microscope. }
    \label{fig:layout}
\end{figure}

The optomechanical resonator used in our experiment is a high-reflectivity Bragg mirror formed of 23 periods of quarter-wave optical thickness GaAs and AlGaAs. The mirror consists of a circular pad with a nominal diameter of $70~\mu$m at the end of a nominally $55~\mu$m long, $8~\mu$m wide, and $220$~nm thick cantilever that is attached to a ledge of the GaAs substrate. 
The fundamental resonance of the mirror is $\Omega_m/2\pi=876$ Hz with a Q-factor of $16,000$ and modal mass of $50$~ng. For more detail on the fabrication of the optomechanical resonator, see Refs \cite{Robi2016Stable,Cripe2019measurement,Cripe_thesis,aggarwal2020room}.

The optomechanical resonator is cryogenically cooled to $50$ K and has a transmissivity of $450$~ppm at optical wavelength $\lambda = 1064$ nm. This differs from the room temperature transmissivity of $250$~ppm in \cite{Aggarwal_thesis} due to thermorefractive effects in the Bragg mirror. 
The optomechanical resonator, together with a standard highly reflective mirror, forms a 1-cm-long Fabry-Perot cavity that is placed under vacuum at \~{}$10^{-8}$~Torr. 
The cavity has a linewidth of $\kappa/2\pi = 1.77$~MHz. 
$79\%$ of the input light is spatially mode-matched to the cavity, and builds up to $0.25$~W of circulating power. 
The rest of the input light does not match the spatial mode of the cavity, and is promptly reflected. The influence of this mode mismatched part of light is considered in the model (see Methods \ref{app:OMT}). 
The frequency of the input light is blue-detuned from the cavity resonance by a fractional detuning of $\delta_\mathrm{cav}=2\Delta/\kappa=0.24$. 
Blue-detuning creates a recoil radiation pressure force on the cantilever (i.e. an optical spring), which self-locks the cavity detuning \cite{Cripe2018Rad}. 
To stabilize the mechanical resonator due to the blue-detuning, we apply feedback to an amplitude modulator before the cavity using the light exiting from the 1-cm mirror, as shown in Fig.~\ref{fig:layout}. 
The optical spring and feedback trap the mechanical oscillator at a frequency of $\Omega_\mathrm{eff}/2\pi=113.7$~kHz. 
Optically stiffening the mechanical oscillator in this way suppresses thermal noise \cite{whittle2021approaching} (see Methods \ref{app:damp}), and pushes the resonance above the frequency band where low frequency technical noises dominate. 

Light exiting the cantilever is directed to a balanced homodyne receiver for phase quadrature detection. 
The resulting photocurrent is digitized for analysis at a 2 MHz sampling rate using a Red Pitaya STEMlab 125-14 for subsequent analysis. 
The overall detection efficiency, $\eta = 0.42$, accounts for optical losses, imperfect homodyne visibility, and the finite quantum efficiency of the photodiodes. 
The combined effects of cavity efficiency, intracavity power, and detection chain yield a high measurement rate of 61~kHz, corresponding to 54\% of the mechanical resonance frequency. This rate is sufficient to violate energy equipartition between the displacement and momentum quadratures.

\section*{Data Analysis and the Conditional Mechanical State}\label{sec:results}

\begin{figure}[t]
    \centering
    \includegraphics[width=1\linewidth]{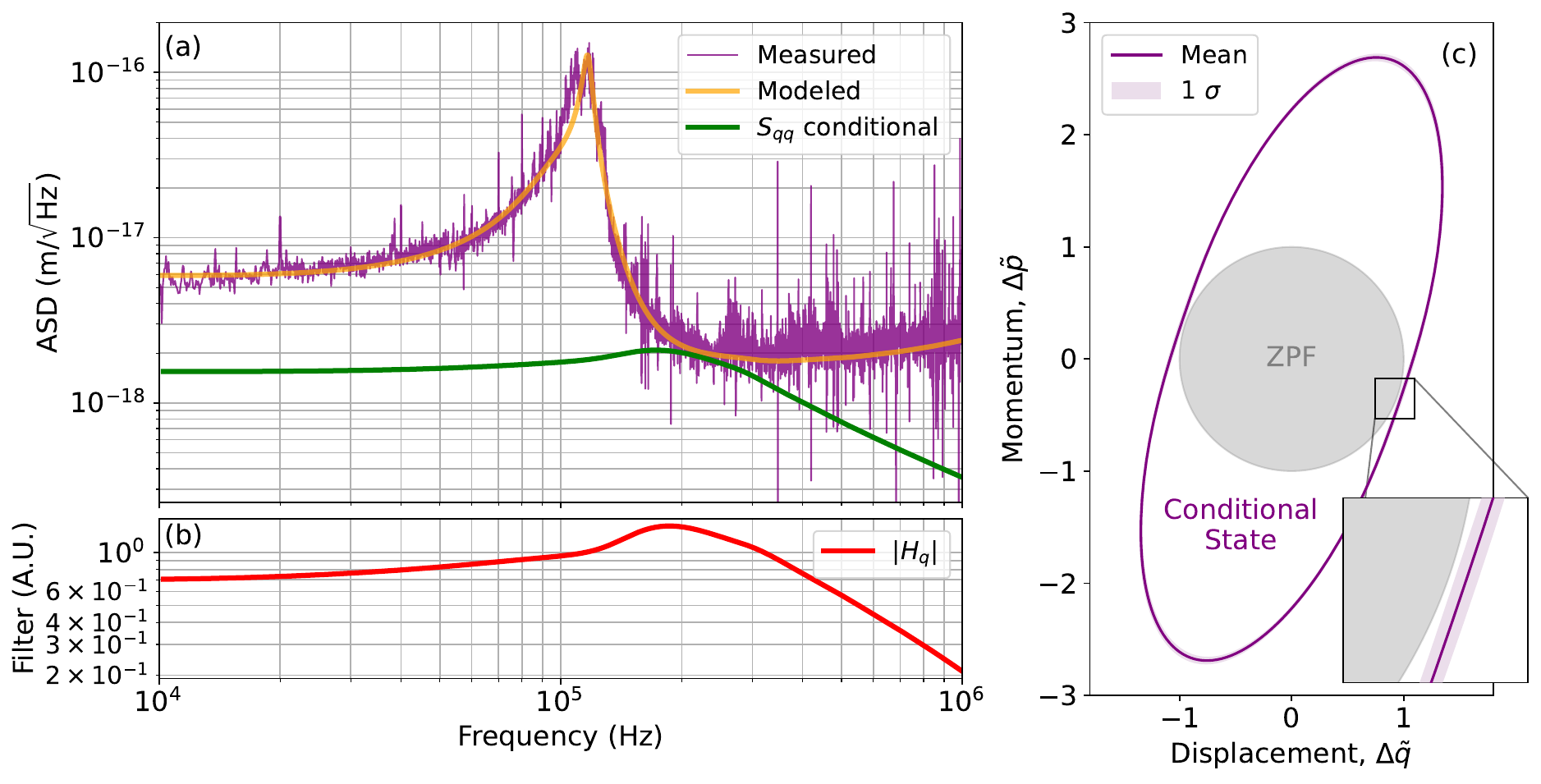}
    \caption{a) Amplitude spectral densities of the directly measured homodyne photocurrent, the calibrated modeled homodyne photocurrent, and the conditional displacement spectrum as in Eqn. \ref{eqn:Sqq}. b) The corresponding displacement Wiener filter as defined in Eqn. \ref{eqn:Hq}. c) Quadrature diagram of the derived conditional state. The light purple band indicates the one-sigma confidence region.}
    \label{fig:main}
\end{figure}

The digitized homodyne photocurrent is saved as a time series. 
We apply a fast Fourier transform to the time series to obtain the voltage spectrum $S_{vv}[\Omega]$,  
which is then calibrated to the displacement spectrum $S_{yy}[\Omega]$ by matching the shot noise to the theoretical model (see Methods \ref{app:calib}). 
We further refine the calibrated spectrum by comparing it with auxiliary spectra measured at different cavity detunings (see Methods \ref{app:noise}), allowing for the removal of noise common to all spectra. 
As shown in Fig.~\ref{fig:main}(a), the cleaned displacement spectrum matches the model $S_{yy}[\Omega]$ well, except in regions affected by non-stationary noise (e.g. frequencies slightly below the mechanical resonance).
To mitigate the influence of non-stationary noise, we use the modeled $S_{yy}[\Omega]$ to compute the causal Wiener filters and to calculate the conditional state. 
We calculate the causal Wiener filters for displacement $\hat{q}$ using the Wiener-Hopf Technique \cite{Kailath1981} as 
\begin{align}
    H_q[\Omega]&=\frac{1}{M_y[\Omega]}\left[\frac{S_{qy}[\Omega]}{M_y[\Omega]^*}\right]_+ \label{eqn:Hq},
\end{align}
where $M_y[\Omega]$ is the causal spectral factor of $S_{yy}[\Omega]$ and $M_y[\Omega]^*$ is the anti-causal spectral factor, such that $M_y[\Omega]M_y[\Omega]^*=S_{yy}[\Omega]$. $[...]_+$ takes the causal decomposition of the function in the bracket (see Chapter 3 of \cite{Kailath1981} for details). 
The cross spectrum $S_{qy}[\Omega]$ is derived directly from the theoretical model. 
$H_q[\Omega]$ is shown in Fig.~\ref{fig:main}(b). 

We then apply the causal Wiener filter to $S_{yy}$ by multiplying them in the frequency domain, yielding the conditional displacement spectrum of the estimation error
\begin{align}
    S_{\delta q\delta q}[\Omega]&=S_{qq}[\Omega]+|H_q[\Omega]|^2 S_{yy}[\Omega]-2\mathrm{Re}[H_q[\Omega]S_{qy}[\Omega]^*] \label{eqn:Sqq},
\end{align}
which is shown in Fig.~\ref{fig:main}(a). 
Similarly, we construct the Wiener filter for $\hat{p}$, and compute the conditional momentum spectrum $S_{\delta p\delta p}$ and cross spectrum $S_{\delta q\delta p}$ (see Methods \ref{app:momentum}). 
As can be seen in Fig.~\ref{fig:main}(a), the conditional spectrum is strongly suppressed compared to the measured and modeled spectra. 
To account for the diverging displacement energy at low frequency given by the structural damping model (see Methods \ref{app:damp} and \cite{meng2022measurement}), we calculate the covariances by integrating the conditional (cross) spectra from $10$~kHz to $1$~MHz to avoid the divergence, e.g. the cross correlation between $\delta\hat{q}$ and $\delta\hat{p}$ is $C_{\delta q\delta p}=\int_{2\pi\times10^4\,\mathrm{Hz}}^{2\pi\times10^6\,\mathrm{Hz}} \frac{d\Omega}{2\pi}\mathrm{Re}[S_{\delta q\delta p}[\Omega]]$. 
Normalizing the (co)-variances to ZPF in displacement $q_\mathrm{zpf}^2=\hbar/2m\Omega_\mathrm{eff}$ and momentum $p_\mathrm{zpf}^2=\hbar m\Omega_\mathrm{eff}/2$, the covariance matrix of the estimation error is 
\begin{align}
    \tilde{\mathbb{V}}=
    \begin{pmatrix}
         \tilde{V}_{\delta q\delta q} & \tilde{C}_{\delta q\delta p}\\
        \tilde{C}_{\delta q\delta p} & \tilde{V}_{\delta p\delta p}
    \end{pmatrix}
    =
    \begin{pmatrix}
        1.76 & 2.08\\
        2.08 & 7.22
    \end{pmatrix},
\end{align}
where the tilde represents normalization to the ZPF. 
The minimum eigenvalue is $1.07\pm0.04$, only $(0.28\pm0.18)$~dB above the ZPF, and the maximum eigenvalue is $7.9\pm0.2$, $(8.98\pm0.11)$~dB above the ZPF level. 
To accurately assess the impact of experimental parameter uncertainties on the conditional variances, the reported mean values and uncertainties are obtained through Monte Carlo uncertainty propagation (see Methods \ref{app:MonteCarlo}).
As a property of Gaussian states, the conditional covariance matrix $\mathbb{V}$, together with the mean values of the estimators $\langle \hat{q}_\mathrm{e} \rangle$ and $\langle \hat{p}_\mathrm{e} \rangle$, uniquely defines the conditional state. The state’s uncertainty is fully characterized by the covariance matrix. 
The quadrature uncertainties corresponding to $\tilde{V}$ are shown in Fig.~\ref{fig:main}(c), with 1 standard deviation uncertainty displayed for reference. 
Our result is the highest quality realization of the protocol to date, reducing the smallest variance by three orders of magnitude compared to the previous record \cite{santiago2020verification}. 

\section*{Discussion and Future Work}\label{sec:discussion}

We have demonstrated the preparation of a classically squeezed mechanical state approaching the quantum regime using a measurement-based protocol. 
The measured spectrum was cleaned by comparing the target spectrum to auxiliary spectra acquired at different cavity detunings. Applying this procedure to the same dataset with various auxiliary spectra confirms the validity of the theoretical model. Although the data cleaning was performed after the measurement was recorded, this approach can be implemented as a real-time filter based on the measured spectra, analogous to causal Wiener filters. Combined with causal Wiener filtering, this enables the prospect of real-time state preparation.

However, our current setup still falls short of preparing a quantum squeezed mechanical motional
state in real-time. 
There are some key further steps required to achieve the goal. First, we need to eliminate non-stationary noise, which prevents direct filtering of the photocurrent for realtime state estimation. Second, a definitive quantum state could be achieved if we could operate the optical cavity closer to resonance, a step that requires a major reworking of the control systems in our setup. Third, we need to improve the readout efficiency, which would be achieved by reducing optical losses throughout the experiment and also increasing the photodetection efficiency. 
Each of these represents a significant technical upgrade in our current setup, and are the subject of future work, with more details provided in Methods \ref{app:impr}. 
With these improvements, the preparation of a quantum-squeezed mechanical state of our 50 ng mirror at room temperature should be achievable \cite{BBLane_thesis}.

Recently, levitated particles have been prepared in quantum-squeezed states in the optical regime~\cite{rossi2024quantum,kamba2025quantum}. Compared to these systems, our optomechanical platform is seven orders of magnitude more massive. This substantial difference in mass enables exploration of a distinct region of the parameter space associated with the quantum-to-classical transition, as well as the potential to probe gravitational effects on quantized motion~\cite{Ludovico2024Testing}.

Even when operating in thermal states and with non-quantum-limited readout, cantilever systems remain among the most promising platforms for investigating the quantum-to-classical transition, provided that suitable materials and structural designs are employed~\cite{Vinante2020Narrowing, carlesso2022present}. Our work advances this effort by enabling significantly more precise tests using potential quantum states and quantum-limited readout.
The steady-state control strategies used in continuous measurement also make it straightforward to adapt these protocols to LIGO test masses.

With feedback stiffening \cite{whittle2021approaching} and ongoing improvements in sensitivity around $100$~Hz \cite{Ganapathy2023Broadband}, preparing the 10-kg-scale LIGO test masses in a quantum-squeezed state should be feasible in the near future. 
Such an achievement would represent a major milestone  in the study of the quantum-to-classical transition. 
Furthermore, LIGO test masses would become the most sensitive quantum-enhanced accelerometers ever developed, directly contributing to dark matter searches \cite{graham2016dark}. 

\backmatter

%\bmhead{Supplementary information}

%If we want supplemental files, we should mention them here.

\bmhead{Acknowledgements}

We wish to thank C. Meng, W. Bowen, V. Sudhir and E. Oelker for fruitful discussions. 
We acknowledge the generous support from the National Science Foundation (NSF). Specifically, J.C. and B.B.L.
are supported through NSF Grant No. PHY-20122088. R.E.P and S.A. are supported by NSF award PHY-2110455. 

\section*{Declarations}

\begin{itemize}
\item Conflict of interest/Competing interests: The authors declare no conflicts of interest. 
\item Data and code availability: Source data for Fig.~\ref{fig:main} is available at \href{https://doi.org/10.7910/DVN/9FGAQF}{https://doi.org/10.7910/DVN/9FGAQF}. Raw data and code underlying the results presented in this paper may be obtained from the authors upon request. 
\item Author contribution: B.B.L. and J.C. constructed the theoretical model, conducted the experiment and analyzed the data; R.E.P., S.A., X.Y. contributed to the preparation of the experiment and data taking; G.D.C. designed and fabricated the cantilever chip; X.Y., T.R.C. and N.M. supervised the project.  
\end{itemize}

\noindent

\begin{appendices}

\section{Optomechanical Model}\label{app:OMT}

Solving the Heisenberg-Langevin equation of a single-mode cavity optomechanical system for the amplitude and phase fluctuations of the intracavity optical field ($\hat{X}$ and $\hat{Y}$, respectively) and the fluctuation in the mechanical displacement $\hat{q}$, we have \cite{rossi2018measurement}
\begin{align}
    \hat{X}[\Omega]=&v[\Omega](\sqrt{2}G\alpha\hat{q}[\Omega]+\sqrt{\kappa_1}\hat{Y}_{\mathrm{in}1}[\Omega]+\sqrt{\kappa_2}\hat{Y}_{\mathrm{in}2}[\Omega]+\sqrt{\kappa_\mathrm{L}}\hat{Y}_{\mathrm{inL}}[\Omega])\nonumber\\
    &+u[\Omega](\sqrt{\kappa_1}\hat{X}_{\mathrm{in}1}[\Omega]+\sqrt{\kappa_2}\hat{X}_{\mathrm{in}2}[\Omega]+\sqrt{\kappa_\mathrm{L}}\hat{X}_{\mathrm{inL}}[\Omega])\\
    \hat{Y}[\Omega]=&u[\Omega](\sqrt{2}G\alpha\hat{q}[\Omega]+\sqrt{\kappa_1}\hat{Y}_{\mathrm{in}1}[\Omega]+\sqrt{\kappa_2}\hat{Y}_{\mathrm{in}2}[\Omega]+\sqrt{\kappa_\mathrm{L}}\hat{Y}_{\mathrm{inL}}[\Omega])\nonumber\\
    &-v[\Omega](\sqrt{\kappa_1}\hat{X}_{\mathrm{in}1}[\Omega]+\sqrt{\kappa_2}\hat{X}_{\mathrm{in}2}[\Omega]+\sqrt{\kappa_\mathrm{L}}\hat{X}_{\mathrm{inL}}[\Omega])\\
    \hat{q}[\Omega]=&\chi_\mathrm{m}[\Omega](\sqrt{2}\hbar G\alpha\hat{X}[\Omega]+\hat{\xi}_\mathrm{th}[\Omega])\label{appeq:q},
\end{align}
where $G=\partial \Omega_\mathrm{cav}/\partial \hat{q}$ is the optomechanical interaction strength, with $\Omega_\mathrm{cav}$ the cavity angular resonance frequency;
$\alpha=\sqrt{n_\mathrm{cav}}$ is the classical amplitude of the intra-cavity optical field, with $n_\mathrm{cav}$ the intra-cavity photon number; 
$\kappa_1$ is the angular cavity coupling rate on the cantilever side, $\kappa_2$ the angular cavity loss rate on the curved mirror side, and $\kappa_\mathrm{L}$ the intra-cavity angular loss rate; 
$\hat{X}_{\mathrm{in}1}$, $\hat{X}_{\mathrm{in}2}$, and $\hat{X}_{\mathrm{inL}}$ are the input amplitude fluctuations from the cantilever, curved mirror and internal loss;
the factors $u[\Omega]=(\kappa/2+i\Omega)/((\kappa/2+i\Omega)^2+\Delta^2)$, $v[\Omega]=-\Delta/((\kappa/2+i\Omega)^2+\Delta^2)$ are the components of the cavity susceptibility, where $\kappa$ is the total angular linewidth of the cavity, and $\Delta$ the angular cavity detuning;
$\hat{\xi}_\mathrm{th}$ is the thermal force operator. 
$\chi_\mathrm{m}[\Omega]=1/(m(\Omega_\mathrm{m}^2-\Omega^2+i\Gamma_\mathrm{m}[\Omega]\Omega))$ is the bare mechanical susceptibility, with $m$, $\Omega_\mathrm{m}$, and $\Gamma_\mathrm{m}[\Omega]$ the effective mass, the natural angular frequency of the mechanical mode, and the natural mechanical linewidth, respectively. 
Here the frequency dependence of $\Gamma_\mathrm{m}$ is a result of structural damping, which will be discussed in detail in Methods \ref{app:damp}. 
In most literature, the optomechanical interaction strength is rewritten in terms of $g=G\alpha x_\mathrm{zpf}$, with $x_\mathrm{zpf}=\sqrt{\hbar/2m\Omega_\mathrm{eff}}$ the ZPF of displacement, where $\Omega_\mathrm{eff}$ is the effective mechanical angular frequency influenced by optical spring and feedback. 
However, we intentionally keep the above form without ZPF, as $\Omega_\mathrm{eff}$ is strongly influenced by optical springing in our system. Keeping $G$ explicitly gives us an expression that works for arbitrary optical spring \cite{Komori2022Quantum}. 

\begin{figure}[h]
    \centering
    \includegraphics[width=0.6\linewidth]{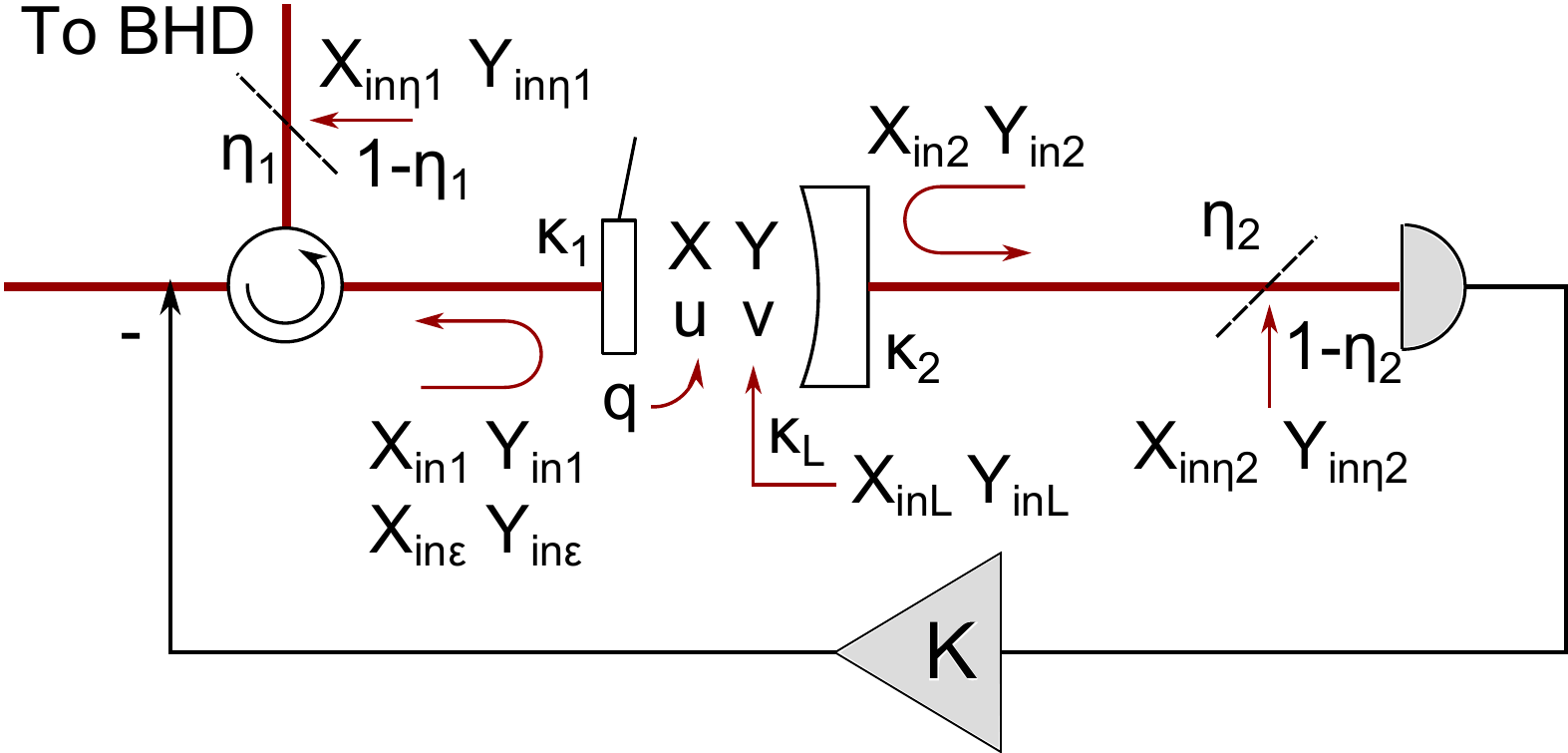}
    \caption{The complete noise model of the experiment. Red paths represent the optical fields, and black paths electronics. }
    \label{fig:nm}
\end{figure}

So far we have worked with an optomechanical system without feedback. 
In practice, as we use an optical spring to stiffen the mechanical response, feedback on the amplitude of the input laser is used to count for the instability caused by the blue detuned laser. 
The feedback loop and various noise are shown in Fig.~\ref{fig:nm}. 

We directly detect the light exiting the curved mirror for feedback control. The open-loop representation of the amplitude fluctuation at the detector is given by
\begin{align}
    \hat{X}_{\mathrm{out}2}[\Omega]=\sqrt{\eta_2}(-\hat{X}_{\mathrm{in}2}[\Omega]+\sqrt{\kappa_2}\hat{X}[\Omega])+\sqrt{1-\eta_2}\hat{X}_{\mathrm{in}\eta2}[\Omega],
\end{align}
where $\eta_2$ is the detection efficiency in the path, and $\hat{X}_{\mathrm{in}\eta2}$ the vacuum fluctuations that leaked into the signal due to imperfect detection. 
The transfer functions of the photo detector, feedback servo, and the amplitude modulator are wrapped up in the controller $K[\Omega]$. 
Then the amplitude modulated light is sent to the optomechanical cavity through the cantilever, and the intra-cavity amplitude fluctuation is given by
\begin{align}
    \hat{X}[\Omega]=&v[\Omega](\sqrt{2}G\alpha\hat{q}[\Omega]+\sqrt{\kappa_1}\hat{Y}_{\mathrm{in}1}[\Omega]+\sqrt{\kappa_2}\hat{Y}_{\mathrm{in}2}[\Omega]+\sqrt{\kappa_\mathrm{L}}\hat{Y}_{\mathrm{inL}}[\Omega])\nonumber\\
    &+u[\Omega](\sqrt{\kappa_1}\hat{X}_{\mathrm{in}1}[\Omega]+\sqrt{\kappa_2}\hat{X}_{\mathrm{in}2}[\Omega]+\sqrt{\kappa_\mathrm{L}}\hat{X}_{\mathrm{inL}}[\Omega])\nonumber\\
    &-u[\Omega]\sqrt{\kappa_1}K[\Omega](\sqrt{\eta_2}(-\hat{X}_{\mathrm{in}2}[\Omega]+\sqrt{\kappa_2}\hat{X}[\Omega])+\sqrt{1-\eta_2}\hat{X}_{\mathrm{in}\eta2}[\Omega]),
\end{align}
where the last term is given by feedback. 
Collecting $\hat{X}[\Omega]$ on both sides, we get the closed-loop expression of $\hat{X}[\Omega]$:
\begin{align}\label{appeq:X}
    \hat{X}[\Omega]=&\frac{1}{1+\sqrt{\eta_2\kappa_1\kappa_2}K[\Omega]u[\Omega]}(v[\Omega](\sqrt{2}G\alpha\hat{q}[\Omega]+\sqrt{\kappa_1}\hat{Y}_{\mathrm{in}1}[\Omega]+\sqrt{\kappa_2}\hat{Y}_{\mathrm{in}2}[\Omega]+\sqrt{\kappa_\mathrm{L}}\hat{Y}_{\mathrm{inL}}[\Omega])\nonumber\\
    &+u[\Omega](\sqrt{\kappa_1}\hat{X}_{\mathrm{in}1}[\Omega]+\sqrt{\kappa_2}\hat{X}_{\mathrm{in}2}[\Omega]+\sqrt{\kappa_\mathrm{L}}\hat{X}_{\mathrm{inL}}[\Omega])\nonumber\\
    &+u[\Omega]\sqrt{\kappa_1}K[\Omega](\sqrt{\eta_2}\hat{X}_{\mathrm{in}2}[\Omega]-\sqrt{1-\eta_2}\hat{X}_{\mathrm{in}\eta2}[\Omega])).
\end{align}
Similarly, the closed-loop $\hat{Y}[\Omega]$ is 
\begin{align}
    \hat{Y}[\Omega]=&u[\Omega](\sqrt{2}G\alpha\hat{q}[\Omega]+\sqrt{\kappa_1}\hat{Y}_{\mathrm{in}1}[\Omega]+\sqrt{\kappa_2}\hat{Y}_{\mathrm{in}2}[\Omega]+\sqrt{\kappa_\mathrm{L}}\hat{Y}_{\mathrm{inL}}[\Omega])\nonumber\\
    &-v[\Omega](\sqrt{\kappa_1}\hat{X}_{\mathrm{in}1}[\Omega]+\sqrt{\kappa_2}\hat{X}_{\mathrm{in}2}[\Omega]+\sqrt{\kappa_\mathrm{L}}\hat{X}_{\mathrm{inL}}[\Omega])\nonumber\\
    &+v[\Omega]\sqrt{\kappa_1}K[\Omega](\sqrt{\eta_2}(-\hat{X}_{\mathrm{in}2}[\Omega]+\sqrt{\kappa_2}\hat{X}[\Omega])+\sqrt{1-\eta_2}\hat{X}_{\mathrm{in}\eta2}[\Omega]),
\end{align}
where we have kept the compact form of $\hat{X}[\Omega]$ in the last line for simplicity. 
Plugging Eq.~\ref{appeq:X} into Eq.~\ref{appeq:q}, and collecting $\hat{q}[\Omega]$, we can see the mechanical susceptibility is modified to be
\begin{align}
    \chi_\mathrm{eff}[\Omega]=\left(\chi_\mathrm{m}^{-1}[\Omega]-\frac{2\hbar G^2 n_\mathrm{cav}v[\Omega]}{1+\sqrt{\eta_2\kappa_1\kappa_2}K[\Omega]u[\Omega]}\right)^{-1}.
\end{align}
The second term in the parenthesis includes the effects of the optical spring in the numerator and feedback in the denominator.  

Now we consider the optical field being sent to the balanced homodyne detector (BHD). 
The input-output relation of a cavity gives an expression of the mode-matched part of the field reflecting off the cantilever $-\hat{X}_{\mathrm{in}1}[\Omega]+\sqrt{\kappa_1}\hat{X}[\Omega]$ for amplitude and $-\hat{Y}_{\mathrm{in}1}[\Omega]+\sqrt{\kappa_1}\hat{Y}[\Omega]$ for phase. 
Due to the non-ideal mode matching of the input laser to the cavity, we have promptly reflected optical field that corresponds to higher order modes of the cavity. 
This field is also amplitude modulated by the feedback, and is given by
\begin{align}
    \hat{X}_{\mathrm{infb}\epsilon}[\Omega]=K[\Omega](\sqrt{\eta_2}(-\hat{X}_{\mathrm{in}2}[\Omega]+\sqrt{\kappa_2}\hat{X}[\Omega])+\sqrt{1-\eta_2}\hat{X}_{\mathrm{in}\eta2}[\Omega])+\hat{X}_{\mathrm{in}\epsilon}[\Omega],
\end{align}
where $\hat{X}_{\mathrm{in}\epsilon}[\Omega]$ is the vacuum fluctuation of this field. 
The phase quadrature is simply $\hat{Y}_{\mathrm{in}\epsilon}[\Omega]$. 
Writing the mode match as $\epsilon$, we have the output amplitude and phase quadratures as
\begin{align}
    \hat{X}_{\mathrm{out}1}[\Omega]=&\sqrt{\eta_1}\left(-e^{i\phi}\sqrt{1-\epsilon}\hat{X}_{\mathrm{infb}\epsilon}[\Omega]+\sqrt{\epsilon}(-\hat{X}_{\mathrm{in}1}[\Omega]+\sqrt{\kappa_1}\hat{X}[\Omega])\right)\nonumber\\
    &+\sqrt{1-\eta_1}\hat{X}_{\mathrm{in}\eta1}[\Omega]\\
    \hat{Y}_{\mathrm{out}1}[\Omega]=&\sqrt{\eta_1}\left(-e^{i\phi}\sqrt{1-\epsilon}\hat{Y}_{\mathrm{in}\epsilon}[\Omega]+\sqrt{\epsilon}(-\hat{Y}_{\mathrm{in}1}[\Omega]+\sqrt{\kappa_1}\hat{Y}[\Omega])\right)\nonumber\\
    &+\sqrt{1-\eta_1}\hat{Y}_{\mathrm{in}\eta1}[\Omega],
\end{align}
where $\eta_1$ is the detection efficiency, $\phi$ the effective Gouy phase difference between the mode-mismatched field and the mode-matched one, and $\hat{X}_{\mathrm{in}\eta1}[\Omega]$ and $\hat{Y}_{\mathrm{in}\eta1}[\Omega]$ the vacuum fluctuations due to loss. 
At the BHD, the detected field is given by
\begin{align}\label{appeq:Xouttheta}
    \hat{X}_{\mathrm{out}1}^\theta[\Omega]=\hat{X}_{\mathrm{out}1}[\Omega]\cos(\theta)+\hat{Y}_{\mathrm{out}1}[\Omega]\sin(\theta),
\end{align}
with $\theta$ the homodyne angle. 

As $\hat{X}_{\mathrm{out}1}[\Omega]$ and $\hat{Y}_{\mathrm{out}1}[\Omega]$ are linear functions of $\hat{q}[\Omega]$, we can write the detected field in displacement unit by dividing both sides of Eq.~\ref{appeq:Xouttheta} by the pre-factor of $\hat{q}$ to get
\begin{align}
    \hat{y}[\Omega]=\hat{q}[\Omega]+\hat{z}[\Omega],
\end{align}
where $\hat{z}$ is the imprecision noise in displacement unit. 
$\hat{y}$ is sometimes referred to as the apparent displacement.

\section{State Estimation and Conditional State}\label{app:StateEst}

State estimation is a broadly used technique to extract a physical observable from a noisy measurement. 
It plays a pivotal role in our work.
Here, we briefly introduce the concepts of state estimation and the conditional state of a mechanical oscillator to lay the foundation for subsequent discussions. 
For any linear continuous measurement of displacement $\hat{q}$, the measurement outcome has the form $\mu\hat{q}+\nu$, with $\mu$ a factor expressing the strength of the measurement and $\nu$ the imprecision noise. 
The measurement outcome can be normalized to displacement units by dividing by $\mu$ as $\hat{y}(t)=\hat{q}(t)+\hat{z}(t)$, where $\hat{z}(t)$ is the normalized measurement imprecision noise.
The measurement process perturbs the mechanical system, and adds noise to $\hat{q}$, which is referred as quantum back-action. 
A typical source of quantum back-action noise is the random momenta transferred from photons in the measurement apparatus to the mechanical oscillator \cite{Aspelmeyer2014Cavity,rossi2018measurement}.
To extract $\hat{q}(t)$ from a noisy measurement record $\hat{y}(t)$, an estimation filter $H_q(t)$ can be constructed and applied using a priori knowledge of the system, such as system parameters and noise spectra \cite{Kailath1981,Wieczorek2015Optimal,Rossi2019Observing}. 
The estimator of displacement is subsequently given by $\hat{q}_\mathrm{e}(t)=H_q*\hat{y}(t)$, with $*$ denoting convolution.
Due to contamination of the signal by imprecision noise $\hat{z}$ and back-action noise, the estimator $\hat{q}_\mathrm{e}$ always deviates from the true value $\hat{q}$. 
The error of the estimate can be written as $\delta\hat{q}(t)=\hat{q}(t)-\hat{q}_\mathrm{e}(t)$.
One can similarly define $\hat{p}_\mathrm{e}$ and $\delta\hat{p}$ for momentum $\hat{p}$. 
For a linear system driven by stationary Gaussian noise, it is well known that the Wiener filters minimize the mean square of estimation errors \cite{Kailath1981}. 

For Gaussian states, the mean values of the estimators $\langle\hat{q}_\mathrm{e}\rangle$ and $\langle\hat{p}_\mathrm{e}\rangle$ and the (co)-variances of their corresponding errors $V_{\delta q\delta q}$, $V_{\delta p\delta p}$ and $C_{\delta q\delta p}$ uniquely define a state centered at the mean values in phase space with uncertainties given by the estimation errors \cite{Ebhardt2009Quantum,Meng2020Mechanical,chen2024causal}.
As the estimators and estimation errors depend on the measurement record $\hat{y}$, the state is deemed a conditional state. 
The mean values directly inherit the stochastic nature of the measurement record $\hat{y}(t)$, resulting in a random walk of the center of the conditional state in phase space, within the distribution of a thermal state given by the thermal and back-action baths, as shown in Fig.~\ref{fig:intuition}(a) in the main text. 
In contrast, the variances of the estimation errors decrease deterministically in time and converge to a steady-state value, as more information is extracted from the measurement process \cite{Rossi2019Observing}. 
The steady-state value is determined by the measurement rate $\Gamma_\mathrm{meas}$, the thermal decoherence rate $\Gamma_\mathrm{th}$ and the back-action rate $\Gamma_\mathrm{ba}$ shown in Fig.~\ref{fig:intuition}(b) and the mechanical frequency $\Omega_\mathrm{m}$ shown in Fig.~\ref{fig:intuition}(c). 
A conditional state prepared by causal filters (constructed using past information) can be transferred to an unconditional state through feedback in real time \cite{BoutHand08}. 
On the other hand, due to the requirement of future information, a conditional state prepared by non-causal filters is not accessible in real time, and cannot be transferred to an unconditional state by feedback. 
Moreover, non-causal filters may result in unphysical conditional states, which violates the Heisenberg uncertainty principle \cite{Zhang2017Prediction}. 
Therefore, for measurement-based preparation of physical mechanical states, estimate filters should be causal.
In addition, to minimize the uncertainty, causal Wiener filters are the typical choices of $H_q$ and $H_p$. 
We follow this practice in our work.

\section{Optical Parameter Characterization}\label{app:param}

Accurate characterization of optical parameters such as cavity linewidth, cavity detuning, and intracavity power is crucial for constructing the theoretical model of the experiment. 
Here, we introduce our method to determine these parameters within 3\% uncertainty. 

\begin{figure}[h]
    \centering
    \includegraphics[width=0.45\linewidth]{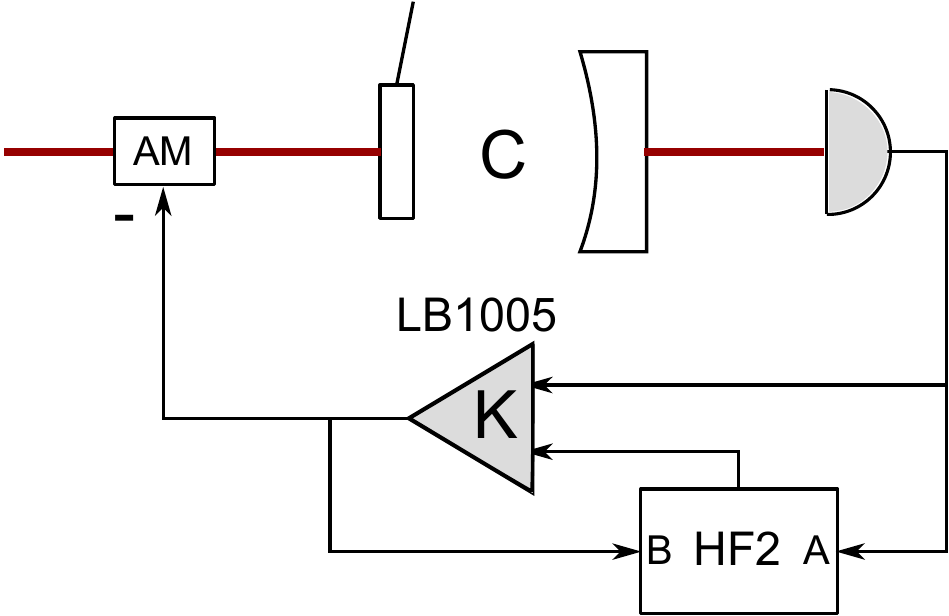}
    \caption{Open-loop cavity transfer function measurement scheme. C and K stand for the transfer functions of the cavity and the control servo (LB1005) respectively. HF2 works as a vector network analyzer. AM stands for the amplitude modulator used to stabilize the cavity.}
    \label{fig:TFmeas}
\end{figure}

We use the optical spring to constrain the optical parameters, as in \cite{Cripe2019measurement}. 
The optical spring frequency is determined by measuring the open-loop cavity transfer function from amplitude fluctuations to photocurrent. 
The measurement scheme is shown in Fig.~\ref{fig:TFmeas}. 
A Newport LB1005 servo is used to lock the cavity, and as a sum point of signal and drive. 
A Z\"{u}rich Instrument HF2 lock-in amplifier is used as the vector network analyzer. The signals right before and after the servo are fed to the two inputs A and B of the HF2, where the ratio of the two is computed. 
The transfer function of the amplitude modulator is flat in the frequency range we are interested in. Therefore, it only contributes a constant factor to the loop, which is absorbed to the controller $K$ for simplicity. 
The closed-loop inputs to Port A and B of the HF2 are given by
\begin{align}
    A &= \frac{CK}{1+CK}d\\ 
    B &= \frac{K}{1+CK}d,
\end{align}
where $d$ is the drive, and $C$ the open-loop transfer function of the cavity. 
Taking the ratio $A/B$, we can recover the desired transfer function $C$. 

\begin{figure}[h]
    \centering
    \includegraphics[width=1\linewidth]{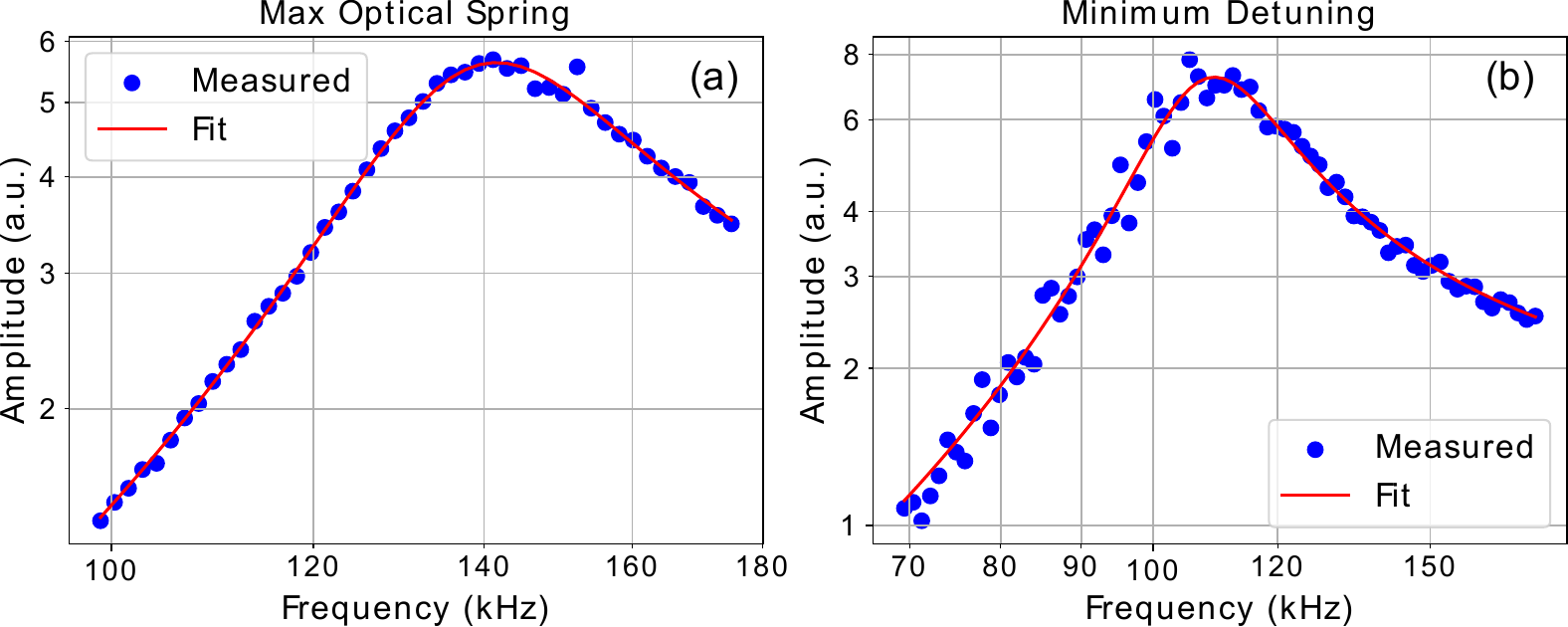}
    \caption{Amplitudes of open-loop cavity transfer functions for (a) maximum optical spring frequency, and (b) minimum cavity detuning.}
    \label{fig:TF}
\end{figure}

Firstly, we lock the cavity on the blue detuning side, and change the cavity length to find the transfer function with maximum optical spring frequency. 
The result is shown in Fig.~\ref{fig:TF}(a). 
In the bad-cavity regime ($\kappa\gg\Omega_\mathrm{eff}$), this corresponds to a fractional detuning of $\delta_\mathrm{cav}=1/\sqrt{3}$.
We fix the detuning and fit the intracavity power and total cavity linewidth. 
The fitting uncertainties of both parameters are below 2\%. 

Secondly, we hold the input power constant, and change cavity length to minimize the detuning. 
As for fitting, we fix the intracavity optical power, which is computed by comparing DC voltages at the photo detector in the maximum optical spring frequency case and the minimum detuning case. 
The intracavity loss is known to change with intracavity power, as a result of radiation pressure changing the alignment of the cavity \cite{Cripe2019measurement}. 
We let the detuning and cavity linewidth to be free parameters. 
The measured transfer function and fitting is shown in Fig.~\ref{fig:TF}(b), and the fitting uncertainties of the parameters are below 3\%. 
As our cavity locking scheme relies on the slope of the cavity resonance, the lowest fractional detuning we can achieve is $\delta_\mathrm{cav}=0.24$. 

As the transfer functions of the servo, amplitude modulator and photo detector are essential for the theoretical model, we also measure them separately using the HF2, and input the results to our model.

\begin{center}
\captionof{table}{Key experimental parameters}
\begin{tabular}{ c|c||c|c } 
 $\kappa/2\pi$ & 1.77~MHz & $\Omega_\mathrm{m}$ & 876~Hz\\ 
 \hline
 $\delta_\mathrm{cav}$ & 0.24 & Q factor & 16000\\ 
 \hline
 $P_\mathrm{cav}$ & 0.25~W & $\Omega_\mathrm{eff}$ & 113.7~Hz\\
 \hline
 $\eta$ & 0.42 & $T$ & 50~K \\
 \hline
 $\lambda$ & 1064~nm & $m$ & 50~ng 
\end{tabular}
\end{center}

\section{Calibration}\label{app:calib}

The homodyne photocurrent power spectrum is measured in units of $\text{V}^2/\text{Hz}$. As the theoretical model is in terms of the cantilever displacement power spectrum with units of $\text{m}^2/\text{Hz}$, we must calibrate our measured spectrum. At small displacements, the cantilever motion induces a linear phase shift on the light reflected from the cavity. The reflected light is then measured using a homodyne detector that is operated within the linear regime, leading to a linear relationship between the measured photocurrent and the cantilever displacement power spectra.

\begin{figure}[t]
    \centering
    \includegraphics[width=0.75\linewidth]{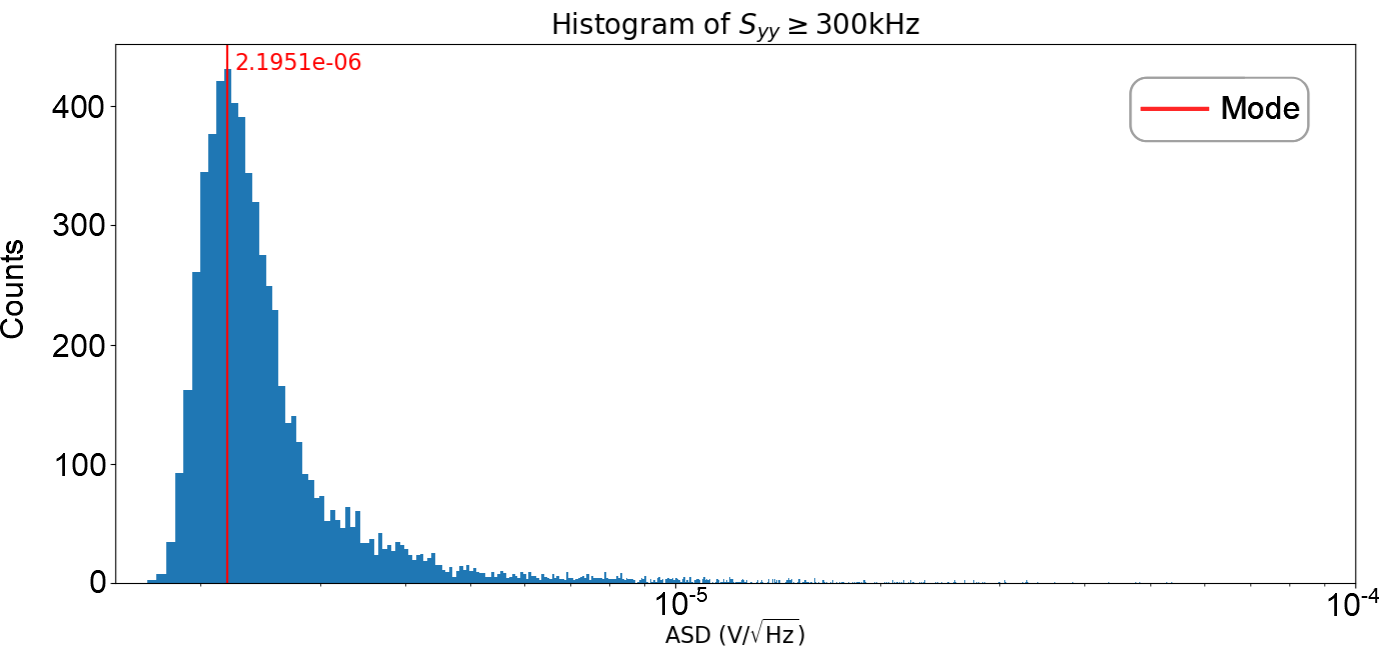}
    \caption{Histogram of the measured homodyne photocurrent amplitude spectral density in the shot noise dominated regime above 300 kHz. The mode of the distribution occurs at an ASD of 2.1951e-6. This is used to calculate the calibration of the raw data from Volts to meters in the presence of excess noise.}
    \label{fig:cal}
\end{figure}

To determine the calibration factor, we align the shot noise dominated regime above $300 \text{kHz}$. As there are narrow regions of excess noise in the raw data, we create a histogram of measured amplitude spectral density (ASD) values as can be seen in Fig. \ref{fig:cal} and use the peak of the distribution for the calibration. With a modeled shot noise of $1.7341\text{e-18}$ $\mathrm{m}/\sqrt{\mathrm{Hz}}$ and a measured shot noise of $2.1951\text{e-6}$ $\mathrm{V}/\sqrt{\mathrm{Hz}}$, the calibration factor is $7.9\text{e-13}$ m/V.

\section{Noisy Background Removal}\label{app:noise}

The direct spectra of the homodyne photocurrents contain noises not included in the theoretical model, such as classical laser phase noise, higher order mechanical modes of the cantilever, mechanical motion of the structure holding the optomechanical cavity, noise from the piezo used to lock the cavity, calibration tones etc. 
Here we introduce a noise removing procedure to recover the spectrum of the fundamental mechanical mode and quantum shot noise from the noisy environment. 
The basic idea is to use homodyne spectra measured with different experimental parameters, especially different cavity detunings, to eliminate the stationary excess noises in the target spectrum. 
In the following, we refer the spectra used to clean the target spectrum as auxiliary spectra. 

As the whole system is linear, the entire experimental apparatus from the optomechanical cavity to the homodyne receiver can be treated as a transfer function from various noises to homodyne photocurrent. 
There are two types of input noises --- force and phase noises. Examples of force noises, which drive actual motion of the mechanical oscillator, include thermal force noise and laser amplitude noise like quantum back-action. 
Phase noises appear as apparent motion due to the sensing apparatus, such as quantum shot noise and classical laser phase noise.
We indicate force noise included in the model as $f_\mathrm{mod}$ and force noise not included as $f_\mathrm{non}$. 
Similarly, we indicate the corresponding phase noises as $\phi_\mathrm{mod}$ and $\phi_\mathrm{non}$ for modeled and unmodeled noises respectively. 
For illustration purpose, we suppose the homodyne receiver detects the intra-cavity phase quadrature. 
Then the force noises are transferred to phase noises through a transfer function $\mathrm{TF}_f$, which is a result of the mechanical susceptibility and the cavity optomechanical interaction. 
Moreover, the cavity and the homodyne transfer phase noise input to photocurrent through $\mathrm{TF}_\phi$. 
We can write the photocurrent of any measurement as 
\begin{align}
    ((f_\mathrm{mod}+f_\mathrm{non})\mathrm{TF}_f+\phi_\mathrm{mod}+\phi_\mathrm{non})\mathrm{TF}_\phi,
\end{align}
and the corresponding spectrum as
\begin{align}
    S_\mathrm{meas}=|((f_\mathrm{mod}+f_\mathrm{non})\mathrm{TF}_f+\phi_\mathrm{mod}+\phi_\mathrm{non})\mathrm{TF}_\phi|^2.
\end{align}
The modeled spectrum in the same case is
\begin{align}\label{appeq:Smod}
    S_\mathrm{mod}=|(f_\mathrm{mod}\mathrm{TF}_f+\phi_\mathrm{mod})\mathrm{TF}_\phi|^2.
\end{align}

Now we consider the auxiliary spectra, formally the same as the expressions above, but with transfer functions and spectra labelled by superscripts ``aux". 
Taking the ratio between $S_\mathrm{meas}^\mathrm{aux}$ and $S_\mathrm{mod}^\mathrm{aux}$, we can eliminate the transfer function $\mathrm{TF}_\phi^\mathrm{aux}$, i.e. the excess noises do not influence how phase noises are transferred into photocurrents: 
\begin{align}
    \frac{S_\mathrm{meas}^\mathrm{aux}}{S_\mathrm{mod}^\mathrm{aux}}=\frac{|(f_\mathrm{mod}+f_\mathrm{non})\mathrm{TF}_f^\mathrm{aux}+\phi_\mathrm{mod}+\phi_\mathrm{non}|^2}{|f_\mathrm{mod}\mathrm{TF}_f^\mathrm{aux}+\phi_\mathrm{mod}|^2}.
\end{align}
For stationary noises, this ratio gives the fraction of the modelled noise to the total noise. 

\begin{figure}[t]
    \centering
    \includegraphics[width=0.75\linewidth]{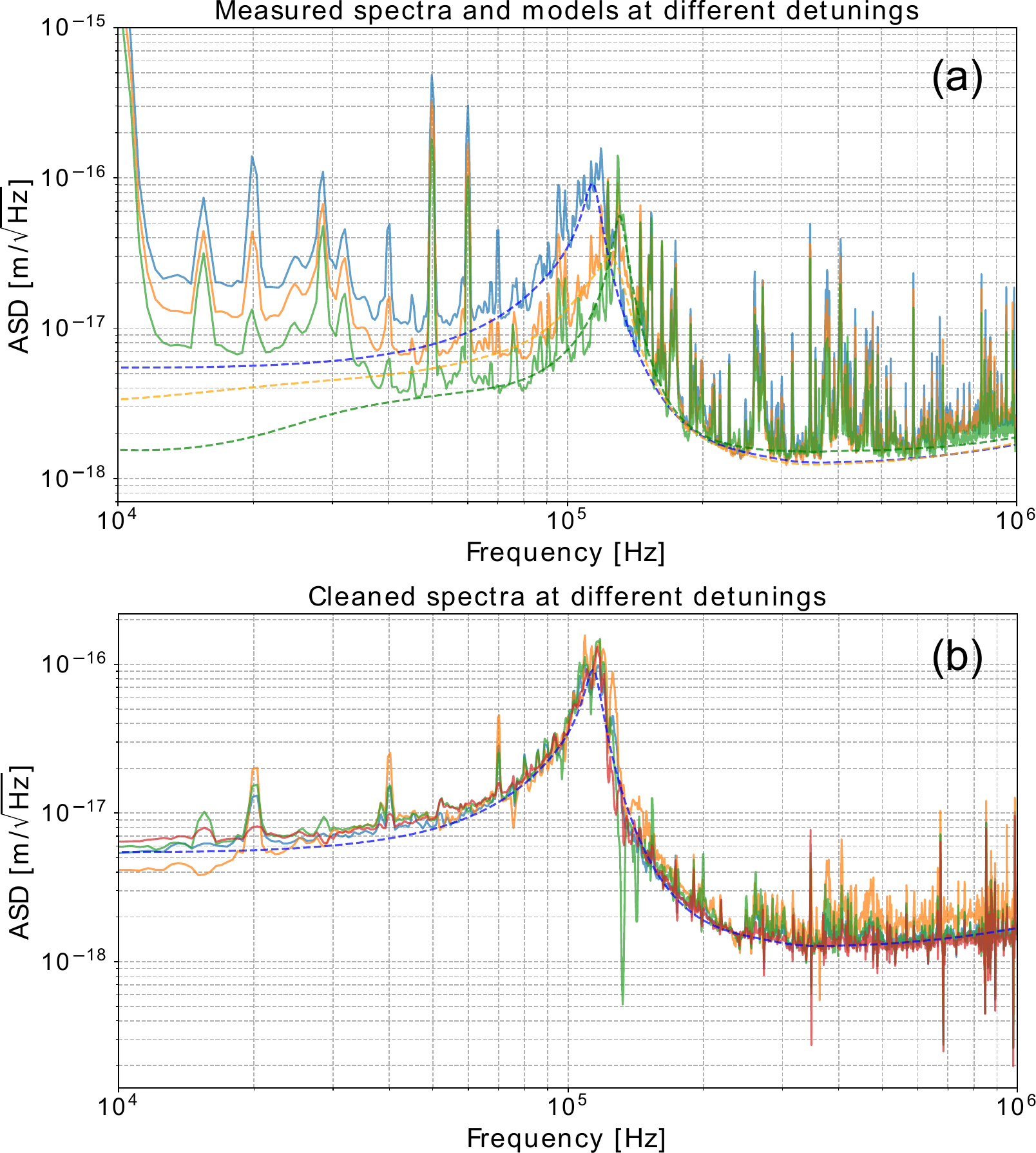}
    \caption{\textbf{(a)} Examples of spectra of homodyne photocurrents (calibrated to displacement unit) at different detunings (solid lines) against the corresponding modelled spectra (dashed lines). Blue, orange and green curves correspond to the lowest, intermediate and the highest detuning used in the measurement respectively. \textbf{(b)} Cleaned spectra of the lowest detuning measurement using 4 different auxiliary measurements. The solid curves are the cleaned data, and the dashed curve is the model. }
    \label{fig:nc}
\end{figure}

Now consider the target noisy measured spectrum $S_\mathrm{meas}^\mathrm{t}$ labelled by superscripts ``t", which is taken with the optimal experimental parameters for mechanical squeezing (with transfer functions $\mathrm{TF}_f^\mathrm{t}$ and $\mathrm{TF}_\phi^\mathrm{t}$), and the same input noise, i.e. under the assumption that the noises are stationary between measurements. We can clean the spectrum by dividing $S_\mathrm{meas}^\mathrm{t}$ by the ratio $S_\mathrm{meas}^\mathrm{aux}/S_\mathrm{mod}^\mathrm{aux}$:
\begin{align}
    S_\mathrm{clean}&=\frac{S_\mathrm{mod}^\mathrm{aux}S_\mathrm{meas}^\mathrm{t} }{S_\mathrm{meas}^\mathrm{aux}}\\
    &=\frac{|f_\mathrm{mod}\mathrm{TF}_f^\mathrm{aux}+\phi_\mathrm{mod}|^2|(f_\mathrm{mod}+f_\mathrm{non})\mathrm{TF}_f^\mathrm{t}+\phi_\mathrm{mod}+\phi_\mathrm{non}|^2 |\mathrm{TF}_\phi^\mathrm{t}|^2}{|(f_\mathrm{mod}+f_\mathrm{non})\mathrm{TF}_f^\mathrm{aux}+\phi_\mathrm{mod}+\phi_\mathrm{non}|^2}. 
\end{align}
Now we evaluate this cleaned data in two regimes: the mechanical motion dominated regime (around and below the stiffened mechanical resonance frequency) and the phase noise dominated regime (at high frequencies). 
In the mechanical motion dominated regime $\phi_\mathrm{mod},\,\phi_\mathrm{non}\rightarrow 0$, we can rewrite the cleaned spectrum as
\begin{align}
    S_\mathrm{clean}&\approx\frac{|f_\mathrm{mod}\mathrm{TF}_f^\mathrm{aux}|^2|(f_\mathrm{mod}+f_\mathrm{non})\mathrm{TF}_f^\mathrm{t}|^2 |\mathrm{TF}_\phi^\mathrm{t}|^2}{|(f_\mathrm{mod}+f_\mathrm{non})\mathrm{TF}_f^\mathrm{aux}|^2}=|f_\mathrm{mod}\mathrm{TF}_f^\mathrm{t}|^2|\mathrm{TF}_\phi^\mathrm{t}|^2\\
    &\approx S_\mathrm{mod}^\mathrm{t}=|(f_\mathrm{mod}\mathrm{TF}_f^\mathrm{t}+\phi_\mathrm{mod})\mathrm{TF}_\phi^\mathrm{t}|^2,
\end{align}
where the modeled spectrum for the target measurement $S_\mathrm{mod}^\mathrm{t}$ inherits the form of Eq.~\ref{appeq:Smod}. 
In the phase noise dominated regime $f_\mathrm{mod},\,f_\mathrm{non}\rightarrow 0$, we have
\begin{align}
    S_\mathrm{clean}\approx\frac{|\phi_\mathrm{mod}|^2|\phi_\mathrm{mod}+\phi_\mathrm{non}|^2 |\mathrm{TF}_\phi^\mathrm{t}|^2}{|\phi_\mathrm{mod}+\phi_\mathrm{non}|^2}=|\phi_\mathrm{mod}|^2|\mathrm{TF}_\phi^\mathrm{t}|^2\approx S_\mathrm{mod}^\mathrm{t}. 
\end{align}
Therefore, in both regimes, the cleaned data recovers the modeled spectrum (i.e. the mechanical motion of the fundamental mode and quantum shot noise) and cleans up the excess noises. 
In general, the cleaned data does not recover the model in the intermediate regime, appearing at the frequencies where the tail of the mechanical peak and the shot noise meet. 
However, experimental data in this regime is covered by loud laser phase noise, making it phase noise dominated again. 
Thus, the data cleaning procedure recovers the motion of the desired mechanical mode and the shot noise from the entire noisy spectrum. 

Figure~\ref{fig:nc} illustrates our noise removal procedure. 
We took homodyne photocurrents for different cavity detunings, the sample spectra of which are shown in Fig.~\ref{fig:nc}(a). The sharp peaks at low frequencies aligning with integer number of 10~kHz are the piezo calibration tone at 10~kHz and its harmonics. At 60~kHz a phase calibration tone overlaps with harmonics of the piezo calibration tone.
At non-integer 10~kHz frequencies, the sharp peaks are high order mechanical modes, which couple weakly to the cavity field, and do not experience significant optical springing. 
The broad structure around 28~kHz corresponds to the cavity piezo resonance. 
The common sharp structures above 130~kHz are phase noises from the laser and fibers. 
Figure~\ref{fig:nc}(b) shows the spectra of the lowest detuning measurement cleaned by 4 different auxiliary spectra. 
All spectra match the modelled curve well. 
We notice that around the optical springed mechanical resonance, there are some noise peaks that are not consistent among the spectra. 
These structures correspond to non-stationary electronic noise. 
The harmonics of the piezo calibration tone are mitigated in the cleaned spectra, but not eliminated, as the nonlinearity of the piezo response depends on the DC bias, which can differ from measurement to measurement. Due to these non-stationary noises, we calculate the final mechanical squeezing using the model instead of the cleaned data.

\section{Momentum and cross spectra of the fundamental mechanical mode}\label{app:momentum}

\begin{figure}[h]
    \centering
    \includegraphics[width=1\linewidth]{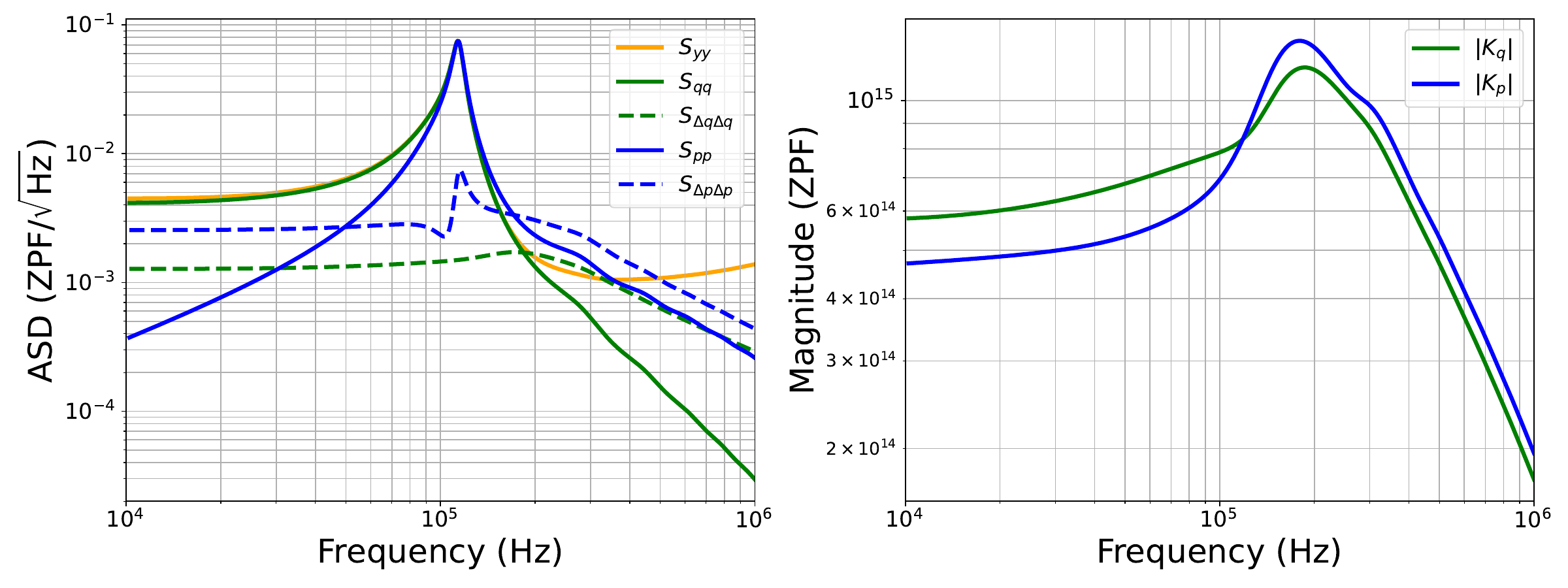}
    \caption{\textbf{Left panel}: ASDs of spectra normalized to ZPF. \textbf{Right panel}: Magnitudes of causal Wiener filters for $\hat{q}$ and $\hat{p}$ normalized to ZPF.}
    \label{fig:q&p}
\end{figure}

In Fig.~\ref{fig:main} of the main text, we only show the displacement related spectra and causal Wiener filter. 
Here we show the momentum related spectra and filter in Fig.~\ref{fig:q&p} for the purpose of completeness. 
The momentum and displacement spectra are related by $S_\mathrm{pp}[\Omega]=m^2\Omega^2 S_\mathrm{qq}[\Omega]$, which is true for both $\hat{q}$, $\hat{p}$ and $\delta\hat{q}$, $\delta\hat{p}$. 
As a result, the momentum spectrum is suppressed at low frequencies, while enhanced at higher frequencies. 

As for the conditional momentum and cross spectra, we construct the causal Wiener filter for $\hat{p}$ similar to $H_q(\Omega)$ as
\begin{align}
    H_p[\Omega]&=\frac{1}{M_y[\Omega]}\left[\frac{S_{py}[\Omega]}{M_y[\Omega]^*}\right]_+,
\end{align}
where $S_{py}$ comes from the model. 
Then the conditional spectra are given by
\begin{align}
    S_{\delta p\delta p}[\Omega]&=S_{pp}[\Omega]+|H_p[\Omega]|^2 S_{yy}[\Omega]-2\mathrm{Re}[H_p[\Omega]S_{py}[\Omega]^*]\\
    S_{\delta q\delta p}[\Omega]&=S_{qp}[\Omega]+H_q[\Omega]H_p[\Omega]^* S_{yy}[\Omega]-H_p[\Omega]^*S_{qy}[\Omega]-H_q[\Omega]S_{py}[\Omega]^*
\end{align}
for momentum and cross spectra respectively.

\section{Structural Damping and Choice of Integration Band}\label{app:damp}

Structural damping is typically the dominant loss channel for mechanical structures in high vacuum \cite{whittle2021approaching,groeblacher2015observation,GonSaul95,Kaj99,NebMav12}. 
The most common model for structural damping is characterized by a constant loss angle $\phi_\mathrm{l}$, which results in the mechanical susceptibility
\begin{equation}\label{chi}
    \chi_\mathrm{m}[\Omega] = \frac{1}{m(\Omega_\mathrm{m}^2-\Omega^2 +i \Omega_\mathrm{m}^2 \phi_\mathrm{l})}=\frac{1}{m(\Omega_\mathrm{m}^2-\Omega^2 +i\Gamma_\mathrm{m}[\Omega]\Omega)},
\end{equation}
where $\Gamma_\mathrm{m}[\Omega]=\Omega_\mathrm{m}^2\phi_\mathrm{l}/\Omega$ has a $1/\Omega$ dependence.
This implies that employing an optical spring to increase the mechanical resonance frequency can improve the thermal coherence of the oscillator \cite{whittle2021approaching}. 
The fluctuation-dissipation theorem gives the thermal displacement spectrum \cite{CalWel51,Kubo66}
\begin{equation}
\begin{split}
    S_{qq}^\mathrm{th}[\Omega] &=-4\hbar\left(n_\mathrm{th}[\Omega]+\frac{1}{2}\right)\mathrm{Im}[\chi_\mathrm{m}[\Omega]] \\
        &= \frac{4\hbar (n_\mathrm{th}[\Omega]+\frac{1}{2})}{m((\Omega^2 -\Omega_\mathrm{m}^2)+(\Omega_\mathrm{m} \phi_\mathrm{l})^2)},
\end{split}
\end{equation}
with $n_\mathrm{th}[\Omega]=1/(\exp[\hbar\Omega/k_\mathrm{B}T]-1)$ being the thermal occupancy of the bath, 
where $k_\mathrm{B}$ is the Boltzmann constant, and $T$ the bath temperature. 
The factor of $4$ in front of $\hbar$ is due to the single-side spectrum. 
In the high temperature limit, $n_\mathrm{th}\approx k_\mathrm{B}T/\hbar\Omega$, which gives $S_{qq}^\mathrm{th}[\Omega]$ a $1/\Omega$ dependence. 
The thermal noise of our cantilever mode reproduces this feature \cite{Cripe2019measurement} as expected. 

However, this model is unphysical. The thermal displacement spectrum diverges at low frequency, which leads to infinite energy and hinders proper state estimation.  
Moreover, the requirement of real time domain susceptibility $\chi_\mathrm{m}(t)$ calls for $\phi_\mathrm{l}[\Omega]$ to be an odd function \cite{Saulson1990Thermal,chen2024causal}. 
A non-zero constant value of $\phi_\mathrm{l}$ certainly violates this rule. 

To avoid this issue, the authors of \cite{meng2022measurement} introduced a roll-off frequency to give $\phi_\mathrm{l}$ a frequency dependence while keep $\int_0^{+\infty}\frac{d\Omega}{2\pi}S_{qq}^\mathrm{th}[\Omega]=(2n_\mathrm{th}+1)V_\mathrm{qzpf}$, i.e. preserve equipartition of thermal noise. 
Alternatively, a Zener model type loss angle can be used instead of a constant $\phi_\mathrm{l}$ \cite{Zen40,chen2024causal}. 
Practically, either of the methods is equivalent to introducing a cut off frequency $\Omega_\mathrm{l}$ to the lower end of the integration band. 
In addition, we can also introduce a cut off frequency $\Omega_\mathrm{h}$ to the higher end for computational convenience. 

We choose cut off frequencies that preserves equipartition of thermal noises, as in \cite{meng2022measurement}. 
However, as the optical spring and feedback shift the mechanical frequency by 2 orders of magnitude, using the natural frequency and linewidth to calculate thermal noise is not a physical choice. 
Instead, we model a mechanical oscillator coupled to the same thermal bath as our real oscillator with susceptibility
\begin{align}
    \chi_\mathrm{m}[\Omega] = \frac{1}{m(\Omega_\mathrm{eff}^2-\Omega^2 +i \Omega_\mathrm{eff}^2 \phi_\mathrm{l})},
\end{align}
where $\Omega_\mathrm{eff}$ is the resonance frequency of the real fundamental mode after optical spring and feedback, and $\phi_\mathrm{l}$ is kept constant. 
Applying the fluctuation dissipation theorem, we can compute $S_\mathrm{qq}^\mathrm{th}$. 
Using the relation $S_\mathrm{pp}^\mathrm{th}[\Omega]=m^2\Omega^2 S_\mathrm{qq}^\mathrm{th}[\Omega]$, we get the spectrum of momentum. 
Then we compute variances normalized to phonon number as
\begin{align}
    V_\mathrm{oo}^\mathrm{th}=\int_{\Omega_\mathrm{l}}^{\Omega_\mathrm{h}}\frac{d\Omega}{2\pi}\frac{S_\mathrm{oo}^\mathrm{th}[\Omega]}{2V_\mathrm{ozpf}},
\end{align}
where $\hat{o}=\hat{q}$ or $\hat{p}$, and the zero-point variance is computed using $\Omega_\mathrm{eff}$. 
We choose $\Omega_\mathrm{h}$ to be as large as possible. As we change $\Omega_\mathrm{l}$ from $10$~Hz to $50$~kHz, the difference $V_\mathrm{qq}^\mathrm{th}/V_\mathrm{pp}^\mathrm{th}-1$ changes from $3^{-4}$ to $2^{-5}$. 
This indicates that the state estimation should be insensitive to the choice of the lower cut off frequency. 
In practice, we choose $\Omega_\mathrm{l}/(2\pi)=10$~kHz, and $\Omega_\mathrm{h}/(2\pi)=1$~MHz for squeezing computation.

\section{Monte Carlo Uncertainty Propagation}\label{app:MonteCarlo}

Our causal Wiener filter and conditional state are calculated using the theoretical model which is based on experimental parameters. 
The uncertainties in these parameters affect the final conditional state. 
Therefore, it is crucial to study the impact of these uncertainties. 
Given the complicated model and filtering process, Monte Carlo uncertainty propagation is the most straightforward way to quantify the impact.

The uncertainties of optical parameters such as $\kappa$, $\delta_\mathrm{cav}$, $P_\mathrm{cav}$ come from the least-square fitting algorithm introduced in Method \ref{app:param}. 
The uncertainty of the temperature is estimated from the temperature range displayed by the thermometer during the experiment. 
We assume that the uncertainties follow Gaussian distributions and pass the mean values and uncertainties of the parameters to the Monte Carlo algorithm, where 985 samples were randomly drawn. 
Due to the non-linearity in the propagation, the means of the covariances of these samples slightly deviates from the values calculated from the parameter means directly. 
We choose to report the Monte Carlo mean. 
The choice does not change the qualitative conclusion of our work.

\section{Future Improvements to Achieve Real-Time Quantum State Preparation}\label{app:impr}

Our current setup falls short of preparing a quantum squeezed mechanical motional state in real-time.
The final steps required to achieve the goal are outlined below. \begin{enumerate}
    \item \textbf{Elimination of Non-Stationary Noise:} \\
    Non-stationary noise in the measurement partially obscures the physical motion of the mechanical oscillator and cannot be fully removed through data cleaning. This noise is observed at multiple locations within the experimental setup and is suspected to originate from electronic sources. Its pervasive nature makes it challenging to identify and eliminate. Currently, recovery of the physical displacement relies on theoretical modeling to compute the causal Wiener filter and the conditional covariance matrix. However, this approach prevents direct filtering of the photocurrent for real-time state estimation. Eliminating non-stationary noise is therefore a critical prerequisite for real-time quantum state preparation.

    \item \textbf{Reducing Cavity Detuning:} \\
    Our results lie at the boundary of the quantum regime. Achieving a definitive quantum-squeezed state can be achieved by reducing the optical spring frequency, which can be realized by locking the optomechanical cavity closer to the resonance, for example using a Pound--Drever--Hall (PDH) locking scheme. The challenges of applying PDH locking scheme to this system remains unknown due to the self-locking effect \cite{Cripe2018Rad}.

    \item \textbf{Enhancing Readout Efficiency :} \\
    Reading out efficiency is another limiting factor of our results, improving which is also a path towards definitive quantum squeezing. A common method is 
    \begin{itemize}
        \item \textit{Optimizing the Optical Injection and
    Detection Configuration}:
    Injecting light from the macroscopic mirror side of the cavity and performing the measurement in transmission can mitigate excess shot noise arising from promptly reflected light that is not spatially matched to the cavity mode. This effectively improves the readout efficiency and helps optimize homodyne visibility.
    \item \textit{Low-loss optics}: readout efficiency can be further improved by optimizing the detection chain and replacing current optical components with lower-loss alternatives.
    \end{itemize}
    
\end{enumerate}

With these improvements, the preparation of a quantum-squeezed mechanical state of a 50~ng mirror at room temperature should be achievable \cite{BBLane_thesis}.

\end{appendices}

%\bibliographystyle{unsrt}
\bibliography{main}% 

%common bib file

\end{document}